\documentclass[twocolumn,epjc3]{svjour3}

\RequirePackage{fix-cm}

\smartqed  
\RequirePackage{graphicx}
\RequirePackage{epstopdf}
\RequirePackage{amssymb,graphics}
\RequirePackage{epsfig,subfigure}
\RequirePackage{color}
\RequirePackage[fleqn]{amsmath}

\newcommand\be{\begin{equation}}
\newcommand\ee{\end{equation}}
\newcommand\ba{\begin{eqnarray}}
\newcommand\ea{\end{eqnarray}}
\newcommand\nn{\nonumber}

\newcommand\pt{\partial}
\journalname{Eur. Phys. J. C}
\begin{document}

\title{Full linear perturbations and localization of gravity on $f(R,T)$ brane}



\author{Bao-Min Gu\thanksref{e1}
        \and Yu-Peng Zhang\thanksref{e2}
         \and Hao Yu\thanksref{e3}
         \and Yu-Xiao Liu\thanksref{e4}
        }

\thankstext{e1}{e-mail: gubm15@lzu.edu.cn}
\thankstext{e2}{e-mail: zhangyupeng14@lzu.edu.cn}
\thankstext{e3}{e-mail: yuh13@lzu.edu.cn}
\thankstext{e4}{e-mail: liuyx@lzu.edu.cn, corresponding author}


\institute{Institute of Theoretical Physics,
            Lanzhou University, Lanzhou 730000,
            People's Republic of China
}

\date{Received: date / Accepted: date}

\maketitle

\begin{abstract}
We study the thick brane world system constructed in the recently proposed $f(R,T)$ theories of gravity, with $R$ the Ricci scalar and $T$ the trace of the energy-momentum tensor. We try to get the analytic background solutions and discuss the full linear perturbations, especially the scalar perturbations. We compare how the brane world model is modified with that of general relativity coupled to a canonical scalar field. It is found that some more interesting background solutions are allowed, and only the scalar perturbation mode is modified. There is no tachyon state exists in this model and only the massless tensor mode can be localized on the brane, which recovers the effective four-dimensional gravity. These conclusions hold provided that two constraints on the original formalism of the action are satisfied.

\end{abstract}
\section{Introduction}

The fundamental idea of brane world~\cite{Akama1982,Visser1985,Arkani-Hamed1998a,Randall1999,Randall1999a,Antoniadis2011} is that the visible universe is localized on a 3-brane which is embedded in a higher-dimensional bulk. A renowned realization is the Randall-Sundrum (RS) brane world model~\cite{Randall1999a}. There, the five-dimensional geometry is a slice of $AdS_5$ due to the negative cosmological constant in the bulk. In such a geometry, the massless graviton is trapped on the brane. As a consequence, the four-dimensional gravity can be recovered even when the extra dimension is infinitely large. This subverts the conventional wisdom that Newton's law means only four non-compact dimensions.

An interesting question is that whether this holds for modified gravity theories. This is the motivation of this paper. In this paper, we consider an RS-like brane world model in modified gravity theory. We use a specific source field instead of the cosmological constant in the original RS model to construct a warped geometry. This actually leads to a domain wall configuration~\cite{DeWolfe2000,Csaki2000a,Gremm2000a,Gremm2000,Kachru:2000hf,Giovannini2001a,Giovannini2002,Kobayashi:2001jd,Sasakura:2002tq,Melfo:2002wd,Bazeia:2003cv,Bazeia:2003aw,Bronnikov:2003gg,Freedman:2003ax,Koley:2005nv,Dzhunushaliev:2009va,Liu2012a,Gu2014}. Note that there is no gravity in the original domain wall model considered by Rubakov and Shaposhnikov~\cite{Rubakov:1983bb}. Usually, in these works one considers a canonical scalar field for simplicity. There are also works where noncanonical scalar field (or K-filed) was considered~\cite{Adam:2007ag,Adam:2008ck,Bazeia2008,ZhongLiu2012,Zhong2014a,Zhong2014},  which would lead to some interesting background solutions and Kaluza-Klein graviton structures (gravity resonances). The models with multiple scalar fields can be seen in Ref.~\cite{Bazeia:2004dh,Aybat:2010sn,George:2011tn}. Recently, an attempt to get a domain wall with an interacting vector field was made in~\cite{GengLu2015}. There, the vector field coupled to gravity nonminimally, resulting a normalizable gravity zero mode.  What is more interesting, it was shown that it is possible to get a domain wall even without both source field and cosmological constant, namely pure geometrical brane~\cite{Arias2002,Barbosa2005kn,Barbosa2006hj,ZhongLiu2015}. These models are based on the modifications of the geometry sector of the Einstein equation, such as $f(R)$ gravity. Inspired by these works, we expect to get a domain wall (thick brane) model in modified gravity theory.

In this paper, we consider the $f(R,T)$ gravity theories~\cite{Harko:2011kv}. This is a special type of modified gravity theory in the sense that it introduces an arbitrary coupling between gravity and source field. As stated in~\cite{Harko:2011kv}, the energy-momentum tensor may not conserve, which implies that the massive particles would not follow the geodesics. In addition, this gravity model has major differences in cosmology and gravitational collapse. Some more works on cosmology with this gravity theory can be seen in Refs.~\cite{Houndjo2011,Jamil2011,Myrzakulov2012,Sharif2012,Sharif2014ioa,Jamil2012,Alvarenga2013,Harko2014,Yousaf2016lls}. The thick brane world model of this theory was considered in Ref.~\cite{Bazeia2015}, and some solutions and the stability of tensor mode were touched.

We will not consider a general $f(R,T)$ in this work. Instead, we will consider a special class of this theory, namely, $f(R,T)=R+F(T)$. We have particular interests in this theory because it can be regarded as a class of the general K-field theory with $L(X,\phi)=K(X,\phi)-V(\phi)$~\cite{ArmendarizPicon1999rj,Garriga1999vw}, which can drive an inflation with general initial data. Here $X$ denotes the kinetic term of the scalar field $\phi$. In Refs. \cite{ZhongLiu2012,Zhong2014a}, the special K-field with $L(X,\phi)=K(X)-V(\phi)$ was used to build domain walls, and some interesting results were obtained. It is straightforward to see that the theory we will consider belong to the $L(X,\phi)=K(X,\phi)-V(\phi)$ type.
To see whether the four dimensional gravity can be recovered, we would like to investigate the full linear gravitational perturbations. The effective four-dimensional gravity can be recovered only when the massless tensor mode is localized on the brane, and the vector and scalar modes are not localized, since the normalizable massless scalar mode or vector mode would lead to a ``fifth force". Except the recovering of four-dimensional gravity, another one of our purposes is to study how the brane world model is modified by replacing $R$ with $R+F(T)$, at both of the background and the perturbation levels. As is well known, the evolutions of perturbations are related to the background configuration, so it is necessary to get the background solutions at first.

We review the $f(R,T)$ gravity theory and its equation of motion in section II, and try to get the background solutions of the brane model with $f(R,T)=R+F(T)$ In section III, we investigate the full linear perturbations with the scalar-vector-tensor decomposition. Then we analyze the behaviors of these perturbation modes, and conclude whether this model gives a viable four-dimensional gravity. At last, we give the conclusions and summary.

\section{Background solutions of the model}\label{solutions}
Let us start with the action and the field equations of the $f(R,T)$ theories of gravity. In five-dimensional spacetime, the action takes the form
 \be
 S=\frac{1}{2\kappa^2}\int d^5 x\sqrt{-g}f(R,T)+ \int d^5 x\sqrt{-g}L_m,
 \label{fRT action}
 \ee
where $f(R,T)$ is an arbitrary smooth function of the Ricci scalar $R$ and of the trace of the stress-energy tensor $T$, and $2\kappa^2=M_{*}^{-3}$ with $M_{*}$ the five-dimensional fundamental scale. The stress-energy tensor is defined by
 \be
 T_{MN}=-\frac{2}{\sqrt{-g}}\frac{\delta(\sqrt{-g}L_m)}{\delta g^{MN}}.
 \label{stress tensor}
 \ee
As discussed in Ref.~\cite{Harko:2011kv}, this formalism of the action allows one to consider a wide class of theories.  In this paper, we consider the choice of $f(R,T)=R+F(T)$, which removes the higher-derivative terms of the field equations.
The gravitational field equation can be obtained by varying the action with respect to the metric $g^{MN}$,  and the result is
\be
R_{MN}-\frac{1}{2}\left(R+F(T)\right)g_{MN}=\kappa^2 T_{MN} + \Theta_{MN},
\label{gravity field equation 1}
\ee
where $\Theta_{MN}=-\frac{\delta F(T)}{\delta g^{MN}}$.
In this work, we consider a canonical scalar field for simplicity, that is
\be
L_m=X-V(\phi)=-\frac{1}{2}\pt_{M}\phi\pt^{M}\phi-V(\phi),
\label{matter lagrangian}
\ee
for which the energy-momentum tensor and its trace are given by
\ba
 T_{MN} &=& {\partial_M\phi } {\partial_N\phi }+ g_{MN} L_m, \\
 T &=&  -\frac{3}{2}g^{MN} {\partial_M\phi } {\partial_N\phi }-5V
  =3X-5V.
\label{matter lagrangian}
\ea
The corresponding equation of motion of the scalar field is
\be
\frac{3}{2\kappa^2}\nabla_{M}\left(F_T\nabla^{M}\phi\right)+\nabla_{M}\nabla^{M}\phi - \left(\frac{5F_T}{2\kappa^2}+1\right)\frac{\pt V(\phi)}{\pt \phi}=0,
\label{scalar field equation}
\ee
here $F_T$ is the derivative of $F(T)$ with respect to $T$. In order to construct a thick brane world model, we use the metric ansatz
\be
ds^2\equiv g_{MN}dx^{M}dx^{N}=e^{2A(y)}\eta_{\mu\nu}dx^{\mu}dx^{\nu}+dy^2.
\label{metric ansatze}
\ee
Here $e^{A(y)}$ is the warp factor, and $x^M=(x^\mu,y)$. With this metric, the field equations~(\ref{gravity field equation 1}) and~(\ref{scalar field equation}) can be expressed in the following specific formalism:
\ba
-6A'^2 - 3A''+\frac{1}{2}F(T)=
\kappa^2 \left(\frac{1}{2}{\phi'}^2+V(\phi)\right),
\label{component field equation 1}\\
-6A'^2 +\frac{1}{2}F(T)=
\kappa^2 \left(-\frac{1}{2}{\phi'}^2+V(\phi)\right)-\frac{3}{2}F_{T}{\phi'}^2,
\label{component field equation 2}
\ea
\ba
\left(\kappa^2+\frac{3}{2}F_T\right)\phi'' + 4\left(\kappa^2+\frac{3}{2}F_T\right)A'\phi'
+\frac{3}{2}F_{T}'\phi'\nonumber
\\
=\left(\kappa^2+\frac{5}{2}F_T\right)V_{\phi},
\label{EOM of scalar field}
\ea
where prime represents the derivative with respect to the extra dimension coordinate $y$. Now the system consists of Eqs.~(\ref{component field equation 1}),~(\ref{component field equation 2}), and (\ref{EOM of scalar field}). There are actually two independent equations because the covariant divergence of the Einstein tensor is zero (the covariant divergence of the energy-momentum tensor does not vanish). We have to solve this system which contains four indeterminate functions, and this implies that we are allowed to impose two constraints on this system.
In the previous work~\cite{Bazeia2015}, a series of solutions including the Sine-Gordon type were obtained by using the first-order formalism equations. The first-order equations were derived by introducing a superpotential, and the equations were solved by giving a specific superpotential. To get more types of solutions except those found in Ref.~\cite{Bazeia2015}, we do not follow this approach in this paper.

\begin{figure}[htb]
\begin{center}
\subfigure[~$\text{e}^{A(y)}$]  {\label{fig1a}
\includegraphics[width=4cm,height=3.0cm]{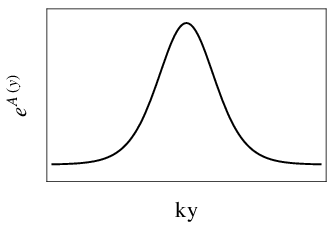}}
\subfigure[~$\phi(y)$]  {\label{fig1b}
\includegraphics[width=4cm,height=3.0cm]{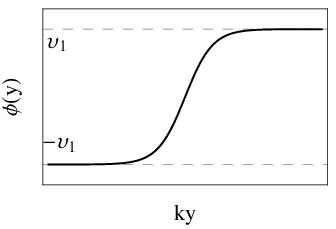}}\vskip 4mm
\subfigure[$V_1 (\phi)$ with $\upsilon_1^2 \kappa ^2\!-\!45>0$]  {\label{fig1c}
\includegraphics[width=4cm,height=3.0cm]{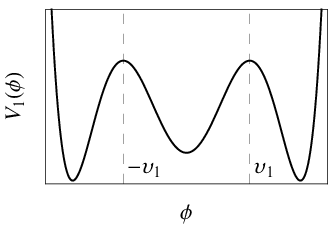}}
\subfigure[$V_1 (\phi)$ with $\upsilon_1^2 \kappa ^2\!-\!45<0$]  {\label{fig1d}
\includegraphics[width=4cm,height=3.0cm]{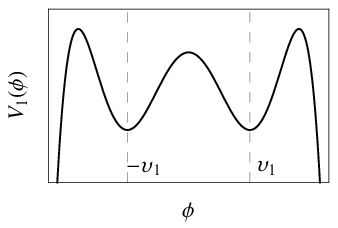}}
\end{center}
\caption{The shapes of the warp factor $e^{A(y)}$ and scalar field $\phi(y)$ with respect to $ky$ and the scalar potential $V_1(\phi)$ with respect to $\phi$ for the first brane solution (\ref{sol warp factor})-(\ref{sol scalar potential}). Note that the  scalar potential is opening up (bottom left) for positive $\upsilon_1^2 \kappa ^2-45$, and opening down (bottom right) for negative $\upsilon_1^2 \kappa ^2-45$.}
\label{fig1}
\end{figure}

As the first example, we consider the simplest case with $F(T)=\alpha T$, for which the effective matter Lagrangian is
\be
L_{eff} = \frac{1}{{2\kappa^2}}F(T)+L_m = (1+\frac{3\alpha}{{2\kappa^2}})X-(1+\frac{5\alpha}{{2\kappa^2}})V.
\label{Leff1}
\ee

Using Eqs.~(\ref{component field equation 1}) and (\ref{component field equation 2}), we get a largely simplified equation
\be
-3A''=\left(\kappa^2+\frac{3}{2}F_T \right)\phi'^2.
\label{munu55 EOM}
\ee
Now the solutions can be obtained by solving Eqs.~(\ref{EOM of scalar field}) and~(\ref{munu55 EOM}). We consider a kink scalar field solution, namely $\phi(y)=\upsilon_1 \text{tanh}(ky)$. This supports the solution of the system as follows
\ba
A(y)&=&
\text{sech}^2(k y)-4 \log \left(\cosh (k y)\right)-1,
\label{sol warp factor}\\
V_{1}(\phi)&=&\frac{9 k^2}{2 \upsilon_1^4 \left(\upsilon_1^2 \kappa ^2-45\right)}
\left(78 \upsilon_1^4 \phi ^2-51 \upsilon_1^2 \phi ^4+8 \phi ^6\right),
\label{sol scalar potential}
\ea
with $\alpha= \frac{12}{\upsilon_1^2}-\frac{2}{3}\kappa ^2$. Note that we have chosen appropriate parameters to make $A(0)=0$. We show the plots of $e^{A(y)}$, $\phi(y)$, and $V(\phi)$ in Figure \ref{fig1}. Clearly, the scalar potential has a $\phi^6$ profile, and it opens up and down for positive and negative $\upsilon_1^2 \kappa ^2-45$, respectively. The scalar field approaches a constant at infinity, which corresponds to the local maxima of the scalar potential for positive $\upsilon_1^2 \kappa ^2-45$. This seems contrary to our common sense. However, it should be noted that the source part of the action and thus the scalar potential are modified. To see this clearly, we investigate the effective Lagrangian (\ref{Leff1}). For constant $F_T$ considered here, we have
\ba
L_{\text{eff}}&=&\frac{18}{c_{0}^{2}\kappa^2}\left(-\frac{1}{2}(\partial\phi)^2 - V_{\text{eff}}(\phi)\right),\\
V_{\text{eff}}(\phi)&=&-\frac{\upsilon_1^2 \kappa^2-45}{27}V(\phi) \nonumber \\
  &=&-\frac{9 k^2}{2 \upsilon_1^4}
\left(78 \upsilon_1^4 \phi ^2-51 \upsilon_1^2 \phi ^4+8 \phi ^6\right).
\ea
From this point of view, the infinity of the extra dimension corresponds to the minimum of the effective scalar potential $V_{\text{eff}}(\phi)$ regardless of the sign of ${c_{0}^{2}\kappa^2} -45$, which is consistent with our conventional wisdom. However, the original scalar potential $V(\phi)$ in (\ref{matter lagrangian}) does not need to follow this due to the inclusion of the $F(T)$ term. For general $F(T)$, there will be some more differences. Now we can see that both of the solutions given by Figs. \ref{fig1c} and \ref{fig1d} are permissible. This is a new feature different from the standard case.

The second example is for a more general power of $T$, i.e., $F(T)=\alpha T^n$ with $n$ a positive integer. The effective Lagrangian is
\be
L_{eff} = \frac{\alpha}{{2\kappa^2}}(3X-5V)^n+X-V.
\label{Leff2}
\ee
One of the solution is given by
\ba
A(y)&=&\log(\text{sech} (k y)),\\
\phi (y)&=&\upsilon_2 \arctan  \left(\tanh  \left({k y}/{2}\right)\right),\\
V_{2}(\phi )&=&\frac{27 k^2}{4 \kappa ^2}\left(\frac{17n-14}{14-9n}-\cos \left({4}\phi/{\upsilon_2} \right)\right), \label{Vcosphi}
\ea
with $\upsilon_2=\frac{6\sqrt{5}}{\kappa }$. It can be easily checked that $T$ is a constant here.  Note that for $n=1$ the action reduces to the first model, but this  solution is different from the first one given in
(\ref{sol warp factor}) and (\ref{sol scalar potential}) since the scalar potential here is the Sine-Gordon one. Similar to the solution in the first example, the infinity of the extra dimension is at the maxima of the scalar potential.

\begin{figure}[htb]
\begin{center}
\subfigure[~$V_2(\phi)$]  {\label{fig2a}
\includegraphics[width=4cm,height=3.0cm]{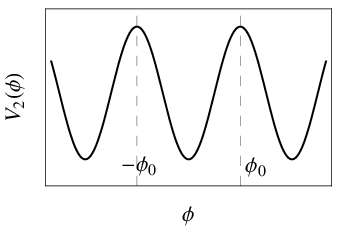}}
\subfigure[~$V_3(\phi)$]  {\label{fig2b}
\includegraphics[width=4cm,height=3.0cm]{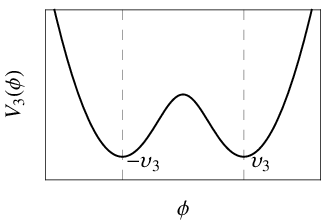}}\vskip 4mm
\end{center}
\caption{The plots of the potentials $V_{2,3}(\phi)$ in the second and third solutions (\ref{Vcosphi}) and (\ref{solution2 V}). The infinity of the extra dimension corresponds to $\pm \phi_0=\pm \frac{\pi \upsilon_2}{4}$ (left) and $\pm \upsilon_3$ (right).}
\label{fig2}
\end{figure}

For nonconstant $F_T$, it is much more difficult to get the solution. We give the solution here without the expression of $F(T)$:
\ba
A (y)&=&\log(\text{sech} (k y)),\label{solution2 a}
\\
\phi(y) &=&\upsilon_3 \tanh  (k y),\label{solution2 phi}
\\
V_{3} (\phi) &=&\frac{9 k^2 }{\upsilon_3^2 \kappa ^2}
\left[ \phi^2\!-\!\frac{15}{2 \kappa ^2} \log
\left(2 \kappa ^2 \left(\phi^2\!-\!\upsilon_3^2\right)
\!+\!15\right)\right], ~~~~~\label{solution2 V}
\\
F(y)&=&k^2 \left(\kappa ^2 \upsilon_3^2 \text{sech}^4(k y)-36 \text{sech}^2(k y)+30\right)
\nonumber
\\
&-&\frac{135 k^2 }{\kappa ^2 \upsilon_3^2}
\log\left(15-2 \kappa ^2 \upsilon_3^2 \text{sech}^2(k y)\right)
.
\label{solution2 F}
\ea
We require that
\ba
\upsilon_3^2<\frac{15}{2 \kappa^2}, \label{Condition v3}
\ea
 to make the log term in (\ref{solution2 F}) to be real.
The plots of the potentials $V_{2}$ and $V_{3}$ are given in Fig.~\ref{fig2}. It can be checked that all the three solutions above give asymptotically $AdS_5$ bulk geometry.

\section{Full Linear Perturbations}

In this section, we discuss the full linear perturbations of this brane world model. From now on, we consider the case of an arbitrary $F(T)$. In the following the calculation is done in the conformally flat coordinate system.  The physical and conformal extra dimension coordinates are related by the equation $dy=e^{A(z)}dz$, which together with (\ref{metric ansatze}) gives
 \be
 ds^2=e^{2A(z)}\left(\eta_{\mu\nu}dx^\mu dx^\nu + dz^2\right).\label{metric in z}
 \ee
Now we introduce the perturbation of this metric. For the background metric $g_{MN}$, the perturbed metric is
 \be
 \tilde{g}_{MN}=g_{MN}+\delta g_{MN}.\label{perturbed metric}
 \ee
Here  $\delta g_{MN}$ is the metric perturbation tensor. To linear order, the metric perturbation can be decomposed into scalar, vector, and tensor modes, or in the following specific formalism
\ba
\!\!\!&&\delta g_{MN}\!=\! e^{2A(z)}\nonumber
\\
\!\!\!&& \times\!
\left(
  \begin{array}{cc}
    \!\!2h_{\mu\nu}\!\!+\!\!\partial_{\mu}\xi_{\nu}\!\!+\!\!\partial_{\nu}\xi_{\mu}\!\!+\!\!2\eta_{\mu\nu}\psi
    \!\!+\!\!2\partial_{\mu}\partial_{\nu}\Phi & ~\zeta_{\mu}\!\!+\!\!\partial_{\mu}\varphi \\
    \zeta_{\nu}\!\!+\!\!\partial_{\nu}\varphi & ~2\chi \\
  \end{array} \!\!
\right).~~~~~
\label{metric perturbation}
\ea
The tensor $h_{\mu\nu}$, which is relevant to the gravitational waves, satisfies the transverse and traceless (TT) condition
\be
\pt^{\mu}h_{\mu\nu}=0,\quad \eta^{\mu\nu}h_{\mu\nu}=0.
\label{TT condition}
\ee
The $\xi_{\nu}$ and $\zeta_{\mu}$ are transverse vector modes, i.e.
\be
\pt^{\mu}\xi_{\mu}=0,\quad \pt^{\mu}\zeta_{\mu}=0.
\ee
The remaining variables $\psi$, $\Phi$, $\varphi$, and $\chi$ represent the scalar degrees of freedom. Clearly, different kinds of modes decouple in the action. As we will see below, this is crucial for our analysis of perturbation modes.

Since we are interested in the behaviours of these modes, it is necessary to get their field equations. The perturbed field equations can be obtained by replacing the background metric in~(\ref{gravity field equation 1}) by the perturbed metric~(\ref{perturbed metric}). If this were done, we would get the field equations that contain the lowest order (zero order), the linear order, and the higher order terms of the metric perturbation. The lowest order parts are just the background equations~(\ref{component field equation 1}) and~(\ref{component field equation 2}). We do not explore the higher order equations since this subject is beyond the scope of the present work. So we concentrate on the linear order field equations. Note that the decomposition~(\ref{metric perturbation}) makes different kinds of modes decouple, therefore we are allowed to divide a linear order equation into three equations (the scalar, vector, and tensor equations).  One can also get the field equations of the metric perturbation modes by writing down the quadratic order action, and then varying this action with respect to various perturbation modes respectively.  We adopt the former path in this work.

The metric perturbation tensor~(\ref{metric perturbation}) contains 7 variables, or 15 degrees of freedom totally. Nevertheless, not all of them are independent because of the gauge invariance. Let us consider the infinitesimal coordinate transformation
\be
x^{A}\rightarrow \tilde{x}^{A}=x^{A} + \epsilon^{A}.
\ee
Under this transformation, the metric perturbation tensor $\delta g_{MM}$ transforms as
\be
\delta g_{MN}\rightarrow \delta \tilde{g}_{MN}=\delta g_{MN} + 2\nabla_{(M}\epsilon_{N)}.
\ee
It is easy to check that the TT part $h_{\mu\nu}$ is gauge invariant. Although the non-TT part is not gauge invariant, it is possible to construct some gauge invariant variables~\cite{Giovannini2001a,Giovannini2002,Kobayashi:2001jd} by using combinations of the variables given in~(\ref{metric perturbation}).
We are allowed to choose a suitable gauge to eliminate the redundant degrees of freedom (5, in five dimensions). Usually, it would be convenient in the longitudinal gauge, i.e. $\xi_{\mu}=0$ and $\Phi=0=\varphi$. In this gauge, the metric perturbation tensor becomes
\be
\delta g_{MN}=e^{2A(z)}
\left(
  \begin{array}{cc}
    2h_{\mu\nu}+2\eta_{\mu\nu}\psi
     & \zeta_{\mu} \\
    \zeta_{\nu} & 2\chi \\
  \end{array}
\right).
\label{metric perturbation 2}
\ee
Using this metric perturbation, we investigate the perturbed field equation of Eq. (\ref{gravity field equation 1}):
\be
\delta R_{MN}-\frac{1}{2}\delta\left(F(T)g_{MN}\right)=\kappa^2 \delta T_{MN}
+\frac{3}{2}\delta\left(F_{T}\partial_{M}\phi\partial_{N}\phi\right).
\label{original perturbation equation}
\ee
To linear order, this equation can be decomposed into $\mu\nu$, $\mu5$, and $55$ components.

The explicit forms of the $\mu\nu$ components are
\ba
\Big[\Box +\pt_{z}^{2}+3\pt_{z}A \pt_{z}
+6\left(\pt_{z}A\right)^2 +2\pt_{z}^{2}A +e^{2A}F  \Big ] h_{\mu\nu}\nn\\
 -2\kappa^2 \left(\frac{1}{2}(\pt_{z}\phi)^2+ e^{2A} V\right)h_{\mu\nu}=0,
\label{tensor equation}
\ea
\be
\partial_{\mu}\pt_{z}\zeta_{\nu}
+\partial_{\nu}\pt_{z}\zeta_{\mu}
+3\pt_{z}A\left(\partial_{\mu}\zeta_{\nu}
+\partial_{\nu}\zeta_{\mu}\right)=0,
\label{vector 0}
\ee
\ba
\!\!\!&&2\eta_{\mu\nu}\Box\psi-2\partial_{\mu}\partial_{\nu}\psi
+3\eta_{\mu\nu}\pt_{z}^{2}\psi+9\eta_{\mu\nu}\pt_{z}A\pt_{z}\psi\nn\\
\!\!\!&&+6\eta_{\mu\nu}\left(\left(\pt_{z}A\right)^2
+\pt_{z}^{2}A\right)\psi-e^{2A}F(T)\eta_{\mu\nu}\psi
\nn\\
\!\!\!&&+\eta_{\mu\nu}\kappa^2\left((\pt_{z}\phi)^2+ 2e^{2A} V(\phi)\right)\psi+\eta_{\mu\nu}\Box\chi-\partial_{\mu}\partial_{\nu}\chi\nn\\
\!\!\!&&-6\eta_{\mu\nu}\left(\left(\pt_{z}A\right)^2
+\pt_{z}^{2}A\right)\chi
-3\eta_{\mu\nu}\pt_{z}A\pt_{z}\chi
\nn\\
\!\!\!&&+\eta_{\mu\nu}\left[\Big(\frac{3}{2}F_T +\kappa^2\Big)\pt_{z}\phi\pt_{z}
+\Big(\frac{5}{2}F_T +\kappa^2\Big)e^{2A}{V_{\phi}}\right]\delta\phi  \nn\\
\!\!\!&&-\eta_{\mu\nu}\Big(\frac{3}{2}F_T +\kappa^2\Big)\left(\pt_{z}\phi\right)^2\chi
=0,
\label{munu component}
\ea
where $\Box=\eta^{\alpha\beta}\partial_\alpha\partial_\beta$ is the four-dimensional d'Alembert operator, and $\delta\phi$ is the perturbation of the scalar field.  Equation~(\ref{tensor equation}) is the equation of motion of the tensor perturbation mode. We can eliminate the first derivative term from~(\ref{tensor equation}) to obtain
\be
\Box\tilde{h}_{\mu\nu}
+\pt_{z}^{2}\tilde{h}_{\mu\nu}
-\left(\frac{3}{2}\pt_{z}^{2}A
+\frac{9}{4}(\pt_{z}A)^{2}\right)\tilde{h}_{\mu\nu}=0
\label{tensor shd equation}
\ee
by defining $\tilde{h}_{\mu\nu}=e^{\frac{3}{2}A}h_{\mu\nu}$. This redefinition is actually equivalent to canonically normalize the kinetic term of the tensor mode.
This is the equation of motion of the tensor mode, and we will analyze it in next section. The remaining two equations~(\ref{vector 0}) and~(\ref{munu component}) lead to
\ba
\pt_{z}\partial_{(\mu}\zeta_{\nu)}
+3\pt_{z}A\partial_{(\mu}\zeta_{\nu)}&=&0,
\label{vector 1}
\\
\chi+2\psi&=&0.
\label{scalar constraint 1}
\ea
As can be seen, Eq.~(\ref{scalar constraint 1}) is just an algebraic equation of the two scalar modes. Usually, one cannot get similar relations in gravity theories with higher-derivative terms of the metric, for instance, the metric formalism $f(R)$ theories of gravity~\cite{DeFelice2010,Sotiriou2010,Nojiri:2010wj}.
Equation~(\ref{vector 1}) is the field equation of the vector mode, and also we will analyze it in next section.

The $\mu5$ components can be divided into two parts
\ba
\kappa^2\left(\frac{1}{2}(\pt_{z}\phi)^2+e^{2A} V\right)\zeta_{\mu}-\frac{1}{2}\Box \zeta_{\mu}
-\frac{1}{2}F(T)e^{2A}\zeta_{\mu}
\nonumber\\
+3\left(\left(\pt_{z}A\right)^2+\pt_{z}^{2}A\right)\zeta_{\mu}=0,
\label{vector 2}
\\
3\pt_{z}A\partial_{\mu}\chi-3\partial_{\mu}\pt_{z}\psi
-\left(\frac{3}{2}F_{T}+\kappa^2\right)\pt_{z}\phi\partial_{\mu}\delta\phi=0.
\label{mu5 component}
\ea
Combining Eqs.~(\ref{component field equation 1}) and~(\ref{vector 2}), we get
\be
\Box\zeta_{\mu}=0.
\label{vector 3}
\ee
This implies that the vector mode is massless. By substituting Eq.~(\ref{scalar constraint 1}) into Eq.~(\ref{mu5 component}), we get the solution of the perturbation of the scalar field
\be
 \delta\phi=-\frac{6(\pt_z \psi + 2\psi\pt_z A )}{(2\kappa^2 + 3 F_T)\pt_z \phi}. \label{delta_phi}
\ee
It should be pointed out that it is impossible to get similar results in gravity theories with higher-derivative terms of source fields, for example, $f(R)$ theories in the Palatini formalism~\cite{Gu2014,Sotiriou2010}. The solution (\ref{delta_phi}) is crucial to the simplification of the $55$ component of Eq.~(\ref{original perturbation equation}), which reads
\ba
\!\!\!&&\pt_{z}\phi\left(\kappa^2+\frac{3}{2}F_{T}
-\frac{9(\pt_{z}\phi)^2}{2e^{2A}}F_{TT}\right)\pt_{z}\delta\phi \!+\! 3\Box\psi\nn\\
\!\!\!&&
\!-e^{2A} {V_{\phi}}\left(\kappa^2 \!+\!\frac{5}{2}F_{T}
 \!+\!
 \frac{15(\pt_{z}\phi)^2}{2e^{2A}} F_{TT}
 \right) \delta\phi
 +12\pt_z A \pt_z\psi
\nn\\
\!\!\!&&=
\left(2\kappa^2 e^{2A} V\!-\!e^{2A}F
 \!-\! \frac{3}{2}F_{T}\left(\pt_{z}\phi\right)^2
 \!-\! \frac{9(\pt_{z}\phi)^4}{2e^{2A}}F_{TT}\right)\chi.~~~~~
\label{55 component}\nn
\ea
Now it is clear that we can reduce the number of scalar perturbation modes of this system to be $1$.
Again using the background equations
(\ref{component field equation 1})--(\ref{EOM of scalar field}) and some manipulations, we get an equation which involves only one scalar mode $\psi$:
\ba
\bigg\{\Box + B(z)\big( \pt_z^2
+ \pt_z \left[\text{ln}\left(\frac{e^{3A}}{G(z)\pt_z\phi}\right)\right]\pt_z ~~~~~~~~~~
\nonumber\\
+ 4\pt_z^2 A- 2\pt_z A \pt_z \text{ln}\left[G(z)\pt_z \phi\right]\big)\bigg\}\psi=0,
\label{scalar 1}
\ea
where $B(z)=1-\frac{9F_{TT}(\pt_z\phi)^3}{2G(z)e^{2A(z)}}$ and $G(z)=\left(\kappa^2 + \frac{3}{2}F_T\right)\pt_z \phi$.
This is the field equation of the scalar mode of the metric perturbation~(\ref{metric perturbation}). Until now we have successfully obtained the field equations of various modes of the metric perturbation~(\ref{metric perturbation}).

\section{Localization of perturbation modes}

In this section let us discuss the behaviors of the tensor, vector, and scalar modes.
This requires the analysis of Eqs.~(\ref{tensor shd equation}),~(\ref{vector 1}) and~(\ref{scalar 1}).

\subsection{Tensor mode}

Equation~(\ref{tensor shd equation}) is a Schr$\ddot{\text{o}}$dinger-like equation of the tensor mode. Clearly, it is the same with that of general relativity. Therefore it has the same mass spectrum. To get a better understanding of the effective four-dimensional gravity, we make a decomposition
\be
\tilde{h}_{\mu\nu}(x^{\mu},z)=\hat{h}_{\mu\nu}(x^{\mu})\Psi(z),
\ee
and then following from~(\ref{tensor shd equation}) we get
\be
\left(\pt_z + \frac{3}{2}\pt_zA\right)\left(-\pt_z + \frac{3}{2}\pt_zA\right)\Psi(z)
=m^2\Psi(z).
\label{factorized shd eq}
\ee
Note that this is consistent with
$\Box \hat{h}_{\mu\nu}=m^2 \hat{h}_{\mu\nu}$. The above equation gives the mass spectrum of the Kaluza-Klein (KK) modes of gravity, and obviously avoids the tachyon instability. The zero mode corresponding to the solution with $m=0$ is given by
\be
\Psi_0 (z)\propto e^{3A(z)/2}.
\label{zero mode}
\ee
The recovering of the effective four-dimensional gravity requires the normalization of the zero mode:
\be
\int |\Psi_0 (z)|^2 dz< \infty.
\label{Tensor Norm condition}
\ee
This is equivalent to have a finite four-dimensional Planck mass, if we define
$M_{\text{Pl}}^2=M_* ^3\int |\Psi_0 (z)|^2 dz$, i.e.
\be
S\supset M_* ^3\int |\Psi_0 (z)|^2 dz \int \partial\hat{h}\partial\hat{h}.
\ee
For our solution~(\ref{sol warp factor}), this condition can surely be satisfied, so the zero mode can be localized.

In addition to the zero mode, Eq.~(\ref{factorized shd eq}) allows a continuous mass spectrum for massive states. If the normalization condition cannot be satisfied for massive state $\Psi_m$, then this massive state would be plane wave at infinity. In other words, the corresponding massive KK graviton cannot be localized on the brane. For our case, all the massive gravitons cannot be localized.

\subsection{Vector mode}
The field equations of the vector mode correspond to Eqs.~(\ref{vector 1}) and~(\ref{vector 3}). They are the same as that of general relativity, so we do not investigate them in detail. It can be concluded straightforwardly that,  if the tensor zero mode can be localized, then the vector mode cannot be localized on the brane.

\subsection{Scalar mode}
We now turn to the scalar mode equation~(\ref{scalar 1}). This equation has significant  difference from that of general relativity~\cite{Giovannini2001a,Giovannini2002,Kobayashi:2001jd}.
To get a Schr$\ddot{\text{o}}$dinger-like formalism equation (or the equation of the canonically normalized field), we first perform a coordinate transformation
\be
dz\equiv \sqrt{B(z)}dr.
\ee
Note that we have a constraint on the function $F(T)$ to make $B(z)>0$.
If this is satisfied then the equation~(\ref{scalar 1}) can be written as
\be
\Box\psi + \pt_r^2 \psi + K(r)\pt_r\psi + J(r)\psi=0,
\label{scalar 2}
\ee
where
\ba
K(r)&=&\pt_r \text{ln}\left(\frac{e^{3A(r)}}{G(r)\pt_r \phi}\right),\\
J(r)&=&4\pt_r^2 A - 2\pt_r A \pt_r\text{ln}[G(r)\pt_r \phi].
\ea

Clearly, since Eq.~(\ref{scalar 2}) contains a term with single derivative on $\psi$, the scalar mode $\psi$ is surely not canonically normalized in the perturbed quadratic action. This trouble can be solved by redefining the scalar mode as $\tilde{\psi}(r)=\text{e}^{\int\frac{1}{2}K(r)dr}\psi(r)$. In terms of $\tilde{\psi}(r)$, the equation~(\ref{scalar 2}) turn to be
\be
\Box\tilde{\psi} + \pt_r^2 \tilde{\psi} + \left[J(r)-\frac{1}{2}\pt_r K-\frac{1}{4}K^2\right]\tilde{\psi}=0.
\ee
The scalar mode can also be separated as $\tilde{\psi}(x^\mu,r)=\hat{\psi}(x^\mu)\Phi(r)$ with $\Box\hat{\psi}(x^\mu)=m^2 \hat{\psi}(x^\mu)$. With this decomposition, we finally get a Schr$\ddot{\text{o}}$dinger-like equation of the canonically normalized  scalar mode $\Phi(r)$:
\be
- \pt_r^2 \Phi + V(r)\Phi=m^2 \Phi,
\ee
with $V(r)=\frac{1}{2}\pt_r K + \frac{1}{4}K^2 - J(r)$. It can be shown that this equation can be factorized as
\be
\left(\pt_r + \frac{\pt_r I(r)}{I(r)}\right)\left(-\pt_r + \frac{\pt_r I(r)}{I(r)}\right)\Phi=m^2 \Phi,
\ee
with $I(r)=\pt_r A [e^{3A}B(r)G(r)\pt_r \phi]^{-1/2}$. To get a real $I(r)$ we require $\kappa^2 + \frac{3}{2}F_T > 0$.
This formalism of equation ensures that no scalar mode with $m^2 <0$ exists.

Furthermore, the zero mode (massless mode) solution is
\be
\Phi_0 \propto I(r).
\ee
As a comparison, we recall the scalar zero mode solution in the standard case, which corresponds to $B(r)=1$ and $G(r)=\kappa^2$.
It is more convenient to analyze the normalization condition in the physical coordinate $y$. Using the background equation~(\ref{munu55 EOM}), we have
\ba
\int \!\! |\Phi_0(r)|^2 dr \!\!&\propto&\!\! \int \!\! \frac{(\pt_z A)^2~dz}
{3e^{3A} \left[(\pt_z A)^2 \!-\! \pt_z^2 A \right] \!- \!9F_{TT} (\partial_z \phi)^4 /2} \nonumber \\
\!\!&\propto& \!\!\int\!\! \frac{(\pt_y A)^2~dy}
{3e^{4A}  \pt_y ^2 A  - 9 e^{3A} F_{TT} (\partial_y \phi)^4 /2}.
\label{normalizationcond}
\ea
It has a significant difference with that of general relativity coupled to a canonical scalar field, in which $F_{TT}=0$.

For our three background solutions given in section~\ref{solutions},  the corresponding $F_{T}$ and $F_{T T}$ are
\ba
F^{(1)}_{T} &=& \alpha=\frac{12}{\upsilon_1^2}-\frac{2}{3}\kappa ^2,\\
F^{(2)}_{T} &=& -\frac{28}{45} \kappa^2 ,\\
F^{(3)}_{T} &=& \frac{2 }{\upsilon_3^2} \cosh ^2(k y)-\frac{2}{3}\kappa ^2,
\ea
and
\ba
F^{(1)}_{T T} &=& 0,\\
F^{(2)}_{T T} &=& -\frac{14(n-1) (9 n-14) \kappa ^4 }{6075 n k^2}~~~(n\ge 1),\\
F^{(3)}_{T T} &=& \frac{\cosh ^6(k y) \left(4 \upsilon_3^2 \kappa ^2-30 \cosh ^2(k y)\right)}{\upsilon_3^4 k^2 \left(6 \upsilon_3^2 \kappa ^2-135 \cosh ^2(k y)\right)},
\ea
respectively. It is easy to verify that all the solutions given in section~\ref{solutions} satisfy the two conditions, $B=1-\frac{9F_{TT}(\pt_y\phi)^2}{2(\kappa^2+\frac{3}{2}F_T)}>0$ and $\kappa^2+\frac{3}{2}F_T >0$. Substituting them to Eq.~(\ref{normalizationcond}), we can see that the scalar zero mode cannot be normalized for all the three solutions.
So our background solutions are stable, and the scalar mode cannot be localized.

It is worth noting here that the first model with $F(T)=\alpha T$ is equivalent to general relativity coupled to a canonical scalar field at background and perturbation levels regardless of whether $T$ is a constant or not. For the second model, namely, $F=\alpha T^n$ with $T$ a constant, even though the action is equivalent to general relativity coupled to a canonical scalar field, they are not equivalent (except $n=1$) at perturbation level. The third model is completely different from the standard case at both of background and perturbation levels.

\section{conclusions}
To summarize, we investigated the thick brane world model in $f(R,T)$ theories of gravity.
The domain wall configuration was constructed by introducing a scalar field in the noncompact bulk. The background solution was obtained by giving a kink scalar field. All of the background quantities are smooth, and so there is no singularity in this asymptotically $AdS_5$ space. In thick brane world models constructed with general relativity coupled to a canonical scalar field, the scalar potential can be $\phi^4$ type and Sine-Gordon type etc, and these solutions share a common characteristic that the vacuum is at the minimum. However, in our model this does not need to be the case. This is a significant new feature that different from the standard case.

Besides, we studied the full linear perturbations of this model, including tensor, vector, and scalar modes. Among these modes, the tensor and vector modes are the same as that of general relativity coupled to a canonical scalar field, and only the scalar mode is modified due to the $F(T)$ term (except for the special case of $F(T)=\alpha T$). We found that, to linear order, the scalar, vector, and tensor modes are stable and no tachyon state exists. Furthermore, we showed that only the tensor zero mode (four-dimensional massless graviton) can be localized on the brane, hence we obtained the viable four-dimensional gravity.  These conclusions hold if the two constraints on the action, namely $B(z)>0$ and $\kappa^2 + \frac{3}{2}F_T > 0$, are satisfied. We argue that these constraints are significant for building viable models from this class of theories.

\section*{Acknowledgements}

This work was supported by the National Natural Science Foundation of China (Grants No. 11522541, and No. 11375075), and the Fundamental Research Funds for the Central Universities (Grant No. lzujbky-2016-k04).

\end{document}